\documentclass[aps,pra,reprint,groupedaddress]{revtex4-1}
\usepackage{graphicx}
\usepackage{amsmath}
\begin{document}


\title{Electronic and magnetic properties of structural defects in SrTiO$_3$(Co)}


\author{I. A. Sluchinskaya}
\author{A. I. Lebedev}
\affiliation{Physics Department, Moscow State University, 119991 Moscow, Russia}


\date{\today}

\begin{abstract}
The synthesis conditions of SrTiO$_3$(Co) samples in which cobalt predominantly
enters the $A$ or $B$~sites of the perovskite $AB$O$_3$ structure are found.
EXAFS studies show that the Co impurity at the $A$~site is off-center and
is displaced from the site by 1.0~{\AA}. XANES studies reveal two predominant
oxidation states of Co: Co$^{2+}$ at the $A$~site and Co$^{3+}$ at the $B$~site.
First-principles calculations of a number of possible
cobalt-containing structural defects reveal defects whose properties are
compatible with the experimentally observed Co oxidation state, its local
structure, magnetic, electrical, and optical properties.

\texttt{DOI: 10.1016/j.jallcom.2019.153243}
\end{abstract}


\maketitle

\section{Introduction}

The discovery of the room-temperature ferromagnetism in cobalt-doped anatase
TiO$_2$~\cite{Science.291.854} after a long period of investigations of
dilute magnetic semiconductors~\cite{JApplPhys.64.R29,Science.281.951} has
initiated a large number of experimental and theoretical studies of wide-gap
oxides doped with magnetic impurities. Later, the room-temperature ferromagnetism
was observed in cobalt-doped ZnO~\cite{PhysRevB.73.024405,ApplPhysLett.79.988},
SnO$_2$~\cite{PhysRevB.77.214436,PhysRevLett.91.077205},
In$_2$O$_3$~\cite{NatureMater.5.298,ApplPhysLett.86.042506},
TiO$_2$~\cite{PhysRevLett.94.157204,ThinSolidFilms.418.197}, and finally
SrTiO$_3$~\cite{ApplPhysLett.89.012501,PhysRevB.84.014416}. The first
experiments were carried out on dilute systems in which the presence of free
carriers was often necessary for the appearance of ferromagnetism. In
Refs.~\cite{NewJPhys.12.043044,PhysRevB.87.144422}, the room-temperature
ferromagnetism was observed in dielectric SrTiO$_3$ samples with high Co
concentration (14--40\%). These materials are promising for their applications
in spintronics, magnetooptics, and photocatalysis. 

The functional properties of magnetic materials can be substantially improved
if another type of ordering, for example, the ferroelectric one, can be
realized in these materials. One of the approaches to creating materials with
such a ``double ordering'' may be doping them with magnetic off-center
impurities. The search for new magnetic off-center impurities, which
simultaneously possess dipole and magnetic moments, is an actual problem
because the interaction of these moments can result in the onset of interesting
magnetoelectric properties~\cite{PhysRevLett.101.165704}. Materials exhibiting
these properties belong to multiferroics---multifunctional materials which open
up new opportunities for modern electronics.

In order to get a high Curie temperature in a dilute ferromagnet, magnetic
atoms should have high magnetic moments. However, the magnetic moments of
magnetic impurities can significantly depend on the structural position and
the local environment of the impurities. Moreover, the impurities of $3d$
transition elements can exist in several oxidation states, which depend on
their structural position, local environment, and on the presence of other
donors and acceptors in the sample.

In practice, the structural position, oxidation and magnetic states can be
changed by varying the synthesis conditions. Our previous studies have shown
that in the case of the Mn impurity in SrTiO$_3$, the distribution of the
impurity between $A$ and $B$~sites of the perovskite $AB$O$_3$ structure and
the impurity oxidation state can be controlled by varying the annealing
temperature and
the stoichiometry of the samples~\cite{JETPLett.89.457,BullRASPhys.74.1235}.
For the Ni impurity in SrTiO$_3$, the magnetic moment was shown to depend on
the structure of a complex formed by the impurity. A nonzero magnetic moment
appears only in the case of the Ni$^{2+}$--$V_{\rm O}$ complex with a distant
oxygen vacancy ($V_{\rm O}$)~\cite{JAdvDielectrics.3.1350031,PhysSolidState.56.449}.

Cobalt-doped strontium titanate has been studied for a long time. The similarity
of crystal structures of SrTiO$_3$ (perovskite) and SrCoO$_{3-x}$ (oxygen-deficient
perovskite) suggests the existence of a continuous series of solid solutions
between them. According to Ref.~\cite{JPowerSources.210.339}, the solubility
of cobalt at the $B$~site of SrTiO$_3$ is at least 40\%, and
Ref.~\cite{InorgChem.43.8169} reported the preparation of single-phase ceramic
samples containing up to 90\% Co by the solid-phase synthesis method. At
$x > 0.5$, X-ray studies~\cite{InorgChem.43.8169} found additional diffraction
peaks, which were attributed to the ordering of the oxygen vacancies. In all
published papers, it was assumed that the Co atom replaces the Ti one. In the
literature there is no data on the incorporation of cobalt into the $A$~site.

To determine the oxidation state of cobalt in SrTiO$_3$, optical absorption, EPR,
X-ray photoelectron spectroscopy (XPS) as well as the study of the near-edge
structure in X-ray absorption spectra (XANES) were used. The optical absorption
spectra of doped single crystals revealed absorption lines characteristic of the
Co$^{3+}$ ion in the octahedral environment~\cite{JPhysC.16.5491,CrystReports.49.469}.
A signal from Co$^{4+}$ ions in the low-spin state ($1\mu_B$) was observed in
EPR studies~\cite{JPhysC.16.5491} of single crystals doped with 0.2\% Co. The
magnitude of this signal significantly increased after the sample illumination
(we guess, in the initial state the Co impurity is diamagnetic and its oxidation
state is +3). In Ref.~\cite{PhysRevB.87.144422}, thin films with Co concentration
from 10 to 50\% were grown by molecular beam epitaxy (MBE) at 550$^\circ$C.
They were studied by XPS,
and a conclusion about the +2 oxidation state of cobalt was made. The same
conclusion was drawn from XPS studies of SrTiO$_3$(Co) nanofibers with Co
concentration of up to 20\% prepared by electrospinning~\cite{JMaterSci.47.8216}
and from XPS studies of thin films containing 30\% Co obtained by pulsed laser
deposition (PLD)~\cite{NewJPhys.12.043044}. The authors of the latter paper
supposed that the difference between oxidation states of Ti and Co is compensated
by the oxygen vacancies and the Co oxidation state in films is lower than in bulk
samples because the films were grown at low partial pressure of oxygen.

It should be noted, however, that the spread of the XPS peaks in different Co
compounds exceeds the magnitude of their systematic chemical shift when changing
the oxidation state~\cite{NIST-XPS-DB}. This is why in
Refs.~\cite{NewJPhys.12.043044,PhysRevB.87.144422} the conclusion about the Co
oxidation state of +2 in SrTiO$_3$ was made from the appearance of a strong
high-energy satellite peak. However, a similar satellite peak with somewhat
lower intensity appears in samples with trivalent cobalt~\cite{PhysRevB.46.9976}.
Therefore, the reliability of this method of determining the oxidation state
raises questions. On the other hand, an indirect argument in favor of the
appearance of Co$^{2+}$ in films obtained by PLD was an increase of the
out-of-plane lattice parameter in doped films~\cite{NewJPhys.12.043044}, which
was in contrast with its decrease in bulk samples~\cite{InorgChem.43.8169,
JMagnMagnMater.305.6}. As for the XANES spectra recorded on SrTiO$_3$(Co)
films~\cite{NewJPhys.12.043044}, no conclusion about the Co oxidation state
was drawn.

The most interesting effect revealed in SrTiO$_3$(Co) is the appearance of
ferromagnetism at room temperature. Studies of magnetic properties have shown
that the ferromagnetism is absent in weakly doped dielectric
samples~\cite{InorgChem.43.8169}. The reason for this can be either the absence
of magnetic moments or their weak interaction. An increase in the Co concentration
to 14--40\% in insulating PLD and MBE films resulted in the appearance of
ferromagnetism~\cite{NewJPhys.12.043044,PhysRevB.87.144422}.
In dilute systems, the distance between the impurities is too large to form
a percolation cluster of strongly interacting magnetic ions. For the appearance
of ferromagnetism in these samples, one can co-dope the samples with donors,
thus creating in them free carriers and conditions for a long-range RKKY
interaction. The study of the influence of La and Nb impurities on the appearance
of ferromagnetism showed the following. In La-doped conductive SrTiO$_3$ PLD
films with Co concentration of 1.5--2\%, the room-temperature ferromagnetism was
observed in as-grown films~\cite{ApplPhysLett.83.2199,PhysRevLett.96.027207}
or in films annealed in a reducing atmosphere~\cite{NewJPhys.11.073042}. In
contrast, in
Nb-doped samples, which became conductive upon doping, no ferromagnetism was
observed down to 5~K in films with 2\% Co grown by PLD at a low oxygen partial
pressure~\cite{ApplPhysLett.89.012501}.

The structure and magnetic properties of SrTi$_{1-x}$Co$_x$O$_3$ epitaxial
films grown by PLD were studied in Ref.~\cite{NewJPhys.12.043044}. In films
with $x = 0.14$ and 0.23, the ferromagnetism was observed at room temperature.
Since these samples were insulating and no exchange via conduction electrons
occurred, the authors supposed that these are the oxygen vacancies which are
responsible for the onset of the ferromagnetism. In the following
paper~\cite{PhysRevB.84.014416}, however, the authors proposed a magnetoelastic
spin ordering model to explain the temperature dependence of magnetization in
the film with $x = 0.23$ up to 1000~K. The important role of the oxygen vacancies
was demonstrated in Ref.~\cite{NewJPhys.11.073042}, in which the ferromagnetism
in epitaxial films appeared only after annealing them in a reducing atmosphere
and disappeared after annealing in an oxidizing atmosphere. It should be
noted, however, that samples annealed in a reducing atmosphere demonstrate
an EPR signal from paramagnetic Ti$^{3+}$ ions~\cite{MaterResBull.16.1149},
whose presence can influence the magnetic properties of reduced samples.
In addition, the contamination of the samples with magnetic impurities can
also affect their properties: a very weak ferromagnetic signal, whose value
was not systematically changed upon reduction, was observed in SrTiO$_3$ samples
free of cobalt at room temperature~\cite{JMagnMagnMater.321.3526}.

A weak ferromagnetism at 300~K was also observed in nanofibers
prepared at $T < 650^\circ$C~\cite{JMaterSci.47.8216}. The magnitude of the
effect strongly depended on the concentration of Co. Moreover, a significant
enhancement of ferromagnetism was observed after annealing the nanofibers in
the hydrogen atmosphere. We note that in this paper the weak ferromagnetism
was observed even in SrTiO$_3$ samples free of cobalt.

In Ref.~\cite{SolidStateElectron.47.2225}, hysteresis loops were observed at
room temperature in samples implanted with Co ions. X-ray studies of these
samples did not reveal precipitates of the second phase. In contrast, in
Ref.~\cite{Ferroelectrics.368.120}, an inhomogeneous depth distribution
of the impurity was observed in Co-implanted samples. It was explained by
the formation of precipitates that can be responsible for the appearance of
ferromagnetism at 300~K.

Single-phase samples of SrTi$_{1-x}$Co$_x$O$_3$ ($x < 0.5$) obtained by the
solid-phase synthesis method at 1250--1400$^\circ$C were studied in
Ref.~\cite{JMagnMagnMater.305.6}. In contrast to the samples described above,
in these samples with $x = {}$0.35--0.50 a transition from the paramagnetic to
the antiferromagnetic state was detected at 15--26~K, and no ferromagnetism was
observed at room temperature. X-ray measurements of the samples reduced in the
10\% H$_2$--Ar atmosphere revealed the appearance of metallic Co. The absence
of the ferromagnetism at $x < 0.5$ and the predominantly antiferromagnetic
ordering of the Co magnetic moments, which resulted from the superexchange
interaction mediated by the oxygen atoms, was also observed in ceramic samples
prepared at 1100$^\circ$C in Ref.~\cite{InorgChem.43.8169}.

In addition to the experimental studies of the room-temperature ferromagnetism,
the theoretical studies that discuss possible mechanisms of this phenomenon
were performed using the HSE06 hybrid functional~\cite{ApplPhysLett.100.252904}
and LDA+$U$ approach~\cite{PhysRevB.87.144422,PhysRevB.90.125130}.
In Ref.~\cite{ApplPhysLett.100.252904},
intermediate spin state ($S = 3/2$) was found for an isolated Co$_B^{4+}$
ion. The ground state for a complex of two next-nearest-neighbor Co ions and
one oxygen vacancy is trivalent Co ions, one of which (adjacent to the vacancy)
is in the high-spin state ($S = 2$) and the other is in the low-spin state
($S = 0$).

An important role of the oxygen vacancies forming the Co$^{2+}_B$--nearest
vacancy complexes whose magnetic moment is $1\mu_B$~\cite{PhysRevB.87.144422}
or even 3$\mu_B$~\cite{PhysRevB.90.125130} (for the same structure) was
demonstrated in LDA+$U$ studies. The choice of divalent cobalt was based on the
conclusion about the Co oxidation state made from the XPS data. Clustering of
Co atoms was found to be energetically unfavorable, in contrast to the
Co$^{2+}_B$--$V_{\rm O}$ association. An isolated Co$^{2+}_B$--$V_{\rm O}$
complex has a magnetic moment of 1$\mu_B$, whereas in the infinite chains of
Co$^{2+}_B$--$V_{\rm O}$ complexes it is 3$\mu_B$. Samples containing both
complexes are insulating. According to Ref.~\cite{PhysRevB.87.144422}, the
magnetic interaction between two nearest-neighbor Co$^{4+}_B$ ions ($S = 1/2$)
is ferromagnetic ($2J = 60$~meV/cell) and becomes very weak for second and third
neighbors. This disagrees with the results of Ref.~\cite{ApplPhysLett.100.252904}
in which the nearest Co atoms are ordered antiferromagnetically and the largest
ferromagnetic interaction was obtained for the next-nearest-neighboring Co$^{4+}_B$
ions. The ferromagnetic interaction of two Co$^{2+}_B$--$V_{\rm O}$ complexes
forming a linear chain is significantly weaker
($2J = {}$8--15~meV/cell)~\cite{PhysRevB.90.125130}.

It should be noted that the properties of the SrTiO$_3$(Co) samples essentially
depend on the method of their preparation. The samples synthesized at high
temperatures (close to thermodynamic equilibrium) are characterized by the presence
of trivalent cobalt and the absence of room-temperature ferromagnetism. In contrast,
in the samples prepared under conditions which can be considered as non-equilibrium
(hydrothermal synthesis, PLD, nanofibers), the cobalt ions become divalent and the
room-temperature ferromagnetism appears. This makes the problem of determining
the real structure of the Co impurity complexes very actual.

In this work, the structural position and the oxidation state of the Co impurity
in SrTiO$_3$ prepared by the solid-phase synthesis method under various
conditions are studied by XAFS (X-ray absorption fine structure) spectroscopy.
It is shown that, depending
on the preparation conditions, cobalt can enter either the $B$~sites of the
perovskite structure in the trivalent state or the $A$~sites in the divalent
state, forming an off-center impurity in the high-spin state ($3\mu_B$).
An analysis of the results of first-principles calculations of the structure
of different Co-containing defects reveals the defects whose properties are
compatible with the experimentally observed Co oxidation state, its local
structure, magnetic, electrical, and optical properties.

\section{Experimental and calculation details}

Cobalt-doped SrTiO$_3$ samples with impurity concentration of 2--3\% and different
deviations from stoichiometry were prepared by the solid-phase synthesis
method. The starting components were SrCO$_3$, nanocrystalline TiO$_2$
obtained by hydrolysis of tetrapropyl orthotitanate and dried at 500$^\circ$C,
and Co(NO$_3$)$_2 \cdot {}$6H$_2$O. The components were weighed in the required
proportions, ground in acetone, and annealed in air in alumina crucibles at
1100$^\circ$C for 4~hours. The obtained powders were ground again and re-annealed
under the same conditions for 4~hours. Some samples were additionally annealed
in air at 1500$^\circ$C or 1600$^\circ$C for 2~hours. To incorporate
cobalt into the $A$ or $B$~sites of the perovskite structure, the composition
of the samples was deliberately deviated from the stoichiometry towards the excess
of titanium or strontium. All obtained samples had a dark brown color.

X-ray absorption spectra in the extended fine structure (EXAFS) and near-edge
structure (XANES) regions were recorded in fluorescence mode at the $K$-edge
of Co (7.709~keV) at 300~K at the KMC-2 station of the BESSY synchrotron radiation
source. The radiation was monochromatized by a (111)-oriented two-crystal
Si$_{1-x}$Ge$_x$ monochromator. The intensity of the incident radiation was
measured using an ionization chamber. The intensity of the fluorescent radiation
was measured by a R{\"O}NTEC silicon drift detector operating in an
energy-dispersive mode. The powders were placed on the surface of adhesive
tape which was then folded to provide an optimal thickness of the sample.

The EXAFS spectra were processed using the widely used IFEFFIT software
package~\cite{IFEFFIT}. The EXAFS function $\chi(k)$ (where
$k = \sqrt{2m(E - E_0)}/\hbar$ is the photoelectron wave vector and $E_0$ is
the absorption edge energy) was extracted from the experimental spectra using
the ATHENA program and fitted using the ARTEMIS program to the theoretical curve
calculated for a given structural model:
    \begin{equation}
    \begin{aligned}
    \chi(k) &= - \frac{1}{k} \sum_j \frac{N_j S_0^2}{R_j^2} |f_j(k)|
    \exp \left (- \frac{2R_j}{\lambda(k)} -2\sigma_j^2 k^2 \right) \\
    &\times \sin(2kR_j + 2\delta_1(k) + \phi_j(k)),
    \label{eq3}
    \end{aligned}
    \end{equation}
where the sum runs over a few nearest shells $j$ of the Co central atom and
$R_j$, $N_j$, and $\sigma_j^2$ are the radius, coordination number, and
Debye--Waller factor for the $j$th shell, respectively. The parameter $S_0^2$
describes the reduction of the oscillation amplitude resulting from
multi-electron and inelastic scattering effects. The backscattering amplitude
$f_j(k)$ and phase shift $\phi_j(k)$ for all single- and multiple-scattering
paths, the phase shift of the central atom $\delta_1 (k)$, and the mean free
path of a photoelectron $\lambda(k)$ were calculated using the FEFF6 program.

For each sample, 3--4 spectra were recorded, they were then independently
processed, and the obtained $\chi$($k$) curves were averaged. The data
processing details are given in Ref.~\cite{PhysRevB.55.14770}.

First-principles calculations of geometry, magnetic and electronic structure
of impurity centers were performed using the ABINIT software package in the
LDA+$U$ approximation. PAW pseudopotentials~\cite{ComputMaterSci.81.446} were
used to describe atoms with a partially filled $d$ shell. The parameters
describing the Coulomb and exchange interaction for Co were $U = 5$~eV and
$J = 0.9$~eV. Our recent study of Ni-doped SrTiO$_3$~\cite{Ferroelectrics.501.1}
has shown that we need supercells having more than 40~atoms to get relatively
narrow impurity bands. That is why to model the properties of isolated Co
centers, we used 80-atom face-centered cubic supercells in which one of the
Ti$^{4+}$ ions at the $B$~site or Sr$^{2+}$
ions at the $A$~site was replaced by the Co ion. To model the properties of
a number of complexes containing two Co atoms, the supercells containing
2$\times$2$\times$3 unit cells (60~atoms) were used. We did not apply the
LDA+$U$ approach to Ti $d$ states because to reproduce the experimental band
gap of SrTiO$_3$, an unphysically high $U$~value of 18.5~eV was needed.

\section{Experimental results}
\label{sec3}

\begin{figure}
\includegraphics{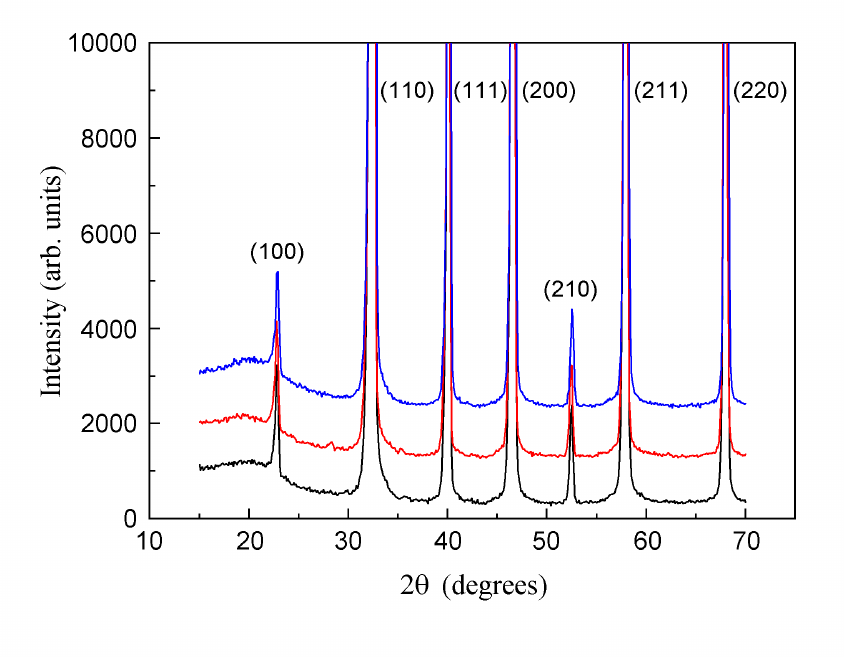}
\caption{X-ray diffraction of SrTiO$_3$(Co) samples: Co2A1600 (black), Co2A1500
(red), and Co3B1100 (blue).}
\label{fig1}
\end{figure}

X-ray diffraction studies of SrTiO$_3$(Co) samples showed that all samples are
single-phase and have the cubic perovskite structure at 300~K (Fig.~\ref{fig1}).
The lattice parameters of the samples were $a = 3.8931(6)$~{\AA} for the Co3B1500
sample,%
    \footnote{The names of the samples include the impurity name, its concentration
    in percents, the site into which the impurity is incorporated, and the annealing
    temperature.}
$a = 3.8994(2)$~{\AA} for the Co2A1500 sample, and $a = 3.8977(5)$~{\AA} for
the Co2A1600 sample. A~decrease in the lattice parameter in doped samples as
compared to that of undoped SrTiO$_3$ ($a = 3.905$~{\AA}) agrees with earlier
published data~\cite{InorgChem.43.8169,JMagnMagnMater.305.6} and indicates the
formation of a solid solution.

\begin{figure}
\includegraphics{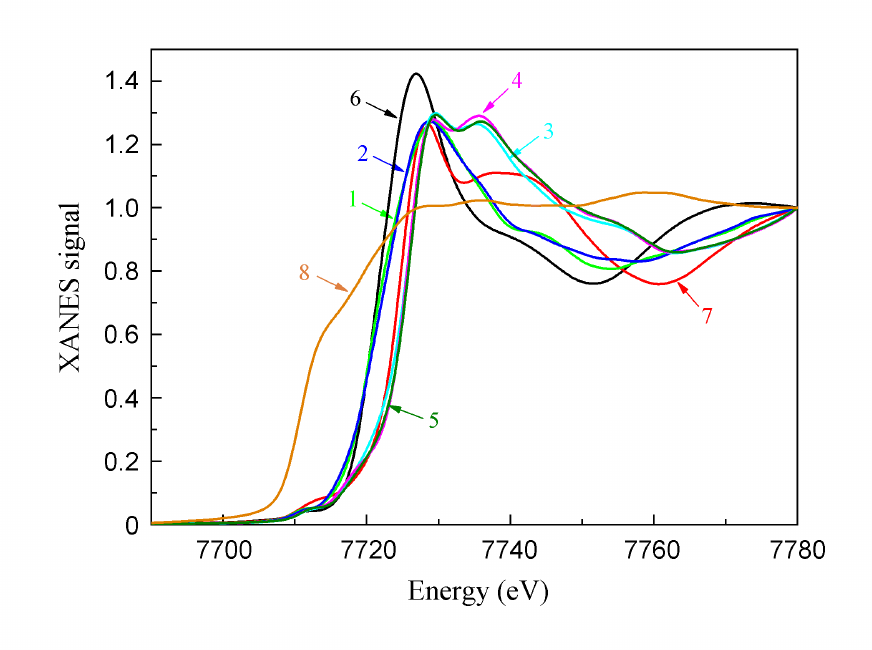}
\caption{XANES spectra of five SrTiO$_3$(Co) samples and three reference
compounds. (1) Co2A1600, (2) Co2A1500, (3) Co2A1100, (4) Co3B1500, (5) Co3B1100,
(6) \mbox{Co(NO$_3$)$_2 \cdot {}$6H$_2$O}, (7) LaCoO$_3$, (8) Co metal foil.}
\label{fig2}
\end{figure}

To determine the oxidation state of the Co impurity in the samples, the position
of the absorption edge in the XANES spectra of doped samples was compared with
those in the Co(NO$_3$)$_2 \cdot {}$6H$_2$O (divalent cobalt) and LaCoO$_3$
(trivalent cobalt) reference compounds (Fig.~\ref{fig2}).

A comparison of the XANES spectra of all samples annealed at 1100$^\circ$C
and also of the sample which had a deviation from stoichiometry toward
an excess of strontium and was annealed at 1500$^\circ$C shows that their
absorption edges
are close to each other and practically coincide with the absorption edge in
the LaCoO$_3$ reference compound. This means that cobalt in these samples is
predominantly in the +3 oxidation state.

The absorption edges of the Sr-deficient SrTiO$_3$(Co) samples annealed at
1500$^\circ$C and 1600$^\circ$C are close to each other and coincide with the
absorption edge in the Co(NO$_3$)$_2 \cdot {}$6H$_2$O reference compound. This
indicates that in these samples cobalt is predominantly in the +2 oxidation
state.

It should be noted that the chemical shift in the XANES spectra is much larger
than that in the XPS spectra~\cite{NIST-XPS-DB}. This allows to draw more
reliable conclusions about the Co oxidation state from XANES spectra.

In order to determine the structural position of the Co impurity, the EXAFS
spectra were analyzed. When analyzing data and choosing structural models,
the lattice parameter obtained from X-ray measurements was taken into account.
In the perovskite structure, the impurity atoms can replace atoms both at the
$A$ and $B$~sites~\cite{JETPLett.89.457}. That is why we considered models with
substitution at both sites.

Comparison of the EXAFS spectra of two groups of samples in which Co atoms
have different oxidation states revealed their qualitative difference.

For all samples in which cobalt is trivalent, a reasonable agreement between
the experimental spectra and theoretically calculated curves was obtained in
the model in which the Co$^{3+}$ ion replaces the Ti$^{4+}$ one with the
formation of a distant oxygen vacancy. In this model, the Co--O distance in
the first shell is 1.909(12)~{\AA}, the Co--Sr distance is 3.342(16)~{\AA},
and the Co--Ti distance is 3.892(12)~{\AA}.

An analysis of the EXAFS spectra of the Co2A1500 and Co2A1600 samples using
the model in which cobalt enters the $B$~site showed that the experimental
and calculated spectra are very different. According to the XANES data,
Co atoms in these samples are mainly in the +2 oxidation state and so they
can replace the $A$~site in SrTiO$_3$. However, there is a big difference
in the ionic radii of Co$^{2+}$ and Sr$^{2+}$ ions. An analysis of the EXAFS
spectra using models in which the Co$^{2+}$ ion replaces the Sr$^{2+}$ ion
showed that a reasonable agreement between the experimental and calculated
curves is obtained in the model in which the Co$^{2+}$ ion is displaced from
the $A$~site along the [100] axis by approximately 1.0~{\AA} (the nearest Co--O
distance is 1.993(42)~{\AA}).

Unfortunately, the agreement between the experimental and calculated EXAFS
curves for all samples was not very good. The agreement criteria were low values
of $R$~factor, good correspondence of obtained coordination numbers to those
of the model, and agreement between Fourier transforms in $R$~space.
In order to understand the reasons for the observed discrepancies, it is
necessary to consider microscopic models of Co impurity centers with atoms
at the $A$ and $B$~sites.

\section{Results of the first-principles calculations}
\label{sec4}

\begin{figure*}
\centering
\includegraphics{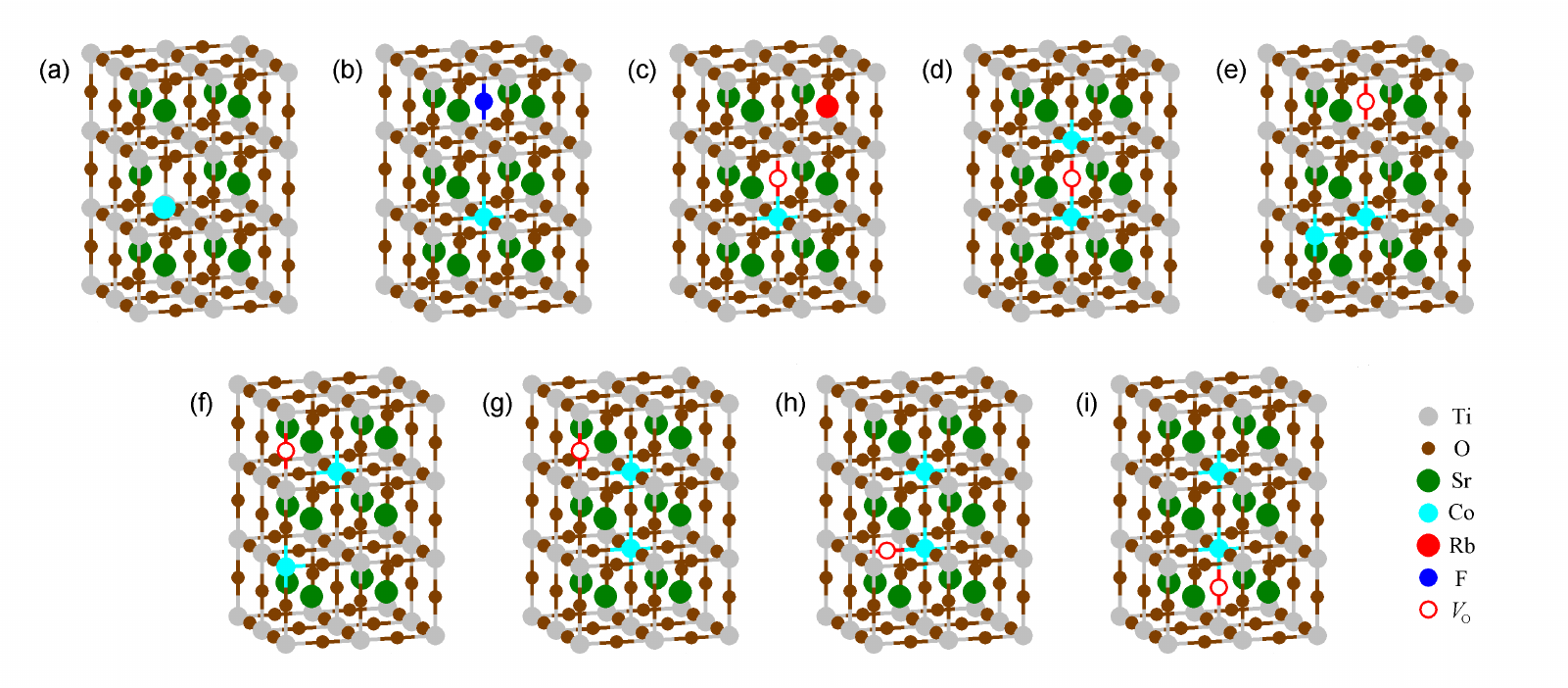}
\caption{\label{fig3}Configurations of different structural models for which
first-principles calculations were performed.}
\end{figure*}

When formulating theoretical models of the impurity centers, the data on the
Co oxidation state and the interatomic distances obtained from the EXAFS data
were taken into account. We note that the earlier first-principles
calculations~\cite{PhysRevB.87.144422,PhysRevB.90.125130,ApplPhysLett.100.252904}
were performed primarily for the Co$^{2+}$ ion at the $B$~site, whereas our
results indicate other oxidation states and other lattice sites.

\begin{table*}
\caption{\label{table1} Averaged distances (in {\AA}) to the neighboring atoms
in the local environment of Co in different models. The distances are corrected
for a systematic underestimation of bond lengths in the LDA approximation.}
\begin{ruledtabular}
\begin{tabular}{lccccccc}
Model & Spin~$S$ & \multicolumn{6}{c}{Shells} \\
\cline{3-8}
      &     & 1 & 2 & 3 & 4 & 5 & 6 \\
\hline
1. Off-center Co$^{2+}_A$ ion, & 3/2 & 2.014\,(4O) & 2.948\,(4Ti) & 3.056\,(1Sr) & 3.168\,(4O) & 3.630\,(4O) & 3.814\,(4O) \\
displacement [100] (Fig.~\ref{fig3}(a)) & \\
2. Off-center Co$^{2+}_A$ ion, & 3/2 & 2.051\,(1O) & 2.069\,(4O) & 2.706\,(2Ti) & 3.265\,(2O) & 3.318\,(2Sr) & 3.474\,(4Ti) \\
displacement [110] & \\
3. Off-center Co$^{2+}_A$ ion, & 3/2 & 1.996\,(3O) & 2.707\,(1Ti) & 2.946\,(6O) & 3.216\,(3Ti) & 3.523\,(3Sr) & 3.564\,(3O) \\
displacement [111] & \\
4. Isolated Co$^{3+}_B$ ion & 0 & 1.905\,(6O) & 3.352\,(8Sr) & 3.801\,(6Ti) & 4.355\,(24O) & 5.509\,(12Ti) \\
(compensated by F) (Fig.~\ref{fig3}(b)) \\
5. Isolated Co$^{3+}_B$ ion & 0 & 1.920\,(6O) & 3.345\,(8Sr) & 3.847\,(6Ti) & 4.366\,(24O) & 5.527\,(12Ti) \\
(electric compensation) \\
6. Co$^{3+}_B$--$V_{\rm O}$ complex & 1 & 1.909\,(5O) & 3.398\,(7Sr,1Rb) & 3.897\,(6Ti) & 4.351\,(24O) & 5.511\,(12Ti) \\
(compensated by Rb) (Fig.~\ref{fig3}(c)) \\
7. Co$^{3+}_B$--$V_{\rm O}$--Co$^{3+}_B$ complex & 1 & 1.895\,(5O) & 3.365\,(8Sr) & 3.897\,(5Ti,1Co) & 4.320\,(24O) & 5.489\,(12Ti) \\
(Fig.~\ref{fig3}(d)) \\
\end{tabular}
\end{ruledtabular}
\end{table*}

First, we calculated the geometry of the impurity center for the Co$^{2+}$ ion
at the $A$~site in SrTiO$_3$. This is a simplest impurity center since the
substitution of Sr$^{2+}$ with Co$^{2+}$ does not require any charge compensation.
Calculations that took into account the full relaxation of atoms showed that
the on-center position of the Co$^{2+}$ ion at the $A$~site is energetically
unstable and the impurity displaces into an off-center position (Fig.~\ref{fig3}(a)).
The equilibrium
displacements of cobalt from the $A$~site were 0.98, 0.94, and 0.75~{\AA} for
[100], [110], and [111] directions, respectively. The comparison of the energies
of the corresponding structures showed that the structure with the [110]
displacement has the lowest energy, whereas the energies of structures with
the [100] and [111] displacements were by 142 and 892~meV higher. The magnetic
moment for all three configurations was $3\mu_B$. The calculated distances to
atoms in the local environment of cobalt for these configurations are given in
Table~\ref{table1}. It is seen that the Co$^{2+}_A$--O distance is notably
larger than the Co$^{3+}_B$--O distance and that the Co$^{2+}_A$--O distances
for [100] and [111] displacements are closest to the experimental value of
1.993(42)~{\AA}.

We start our calculations of impurity centers with Co at the $B$~site with
an isolated Co$^{4+}_B$ ion. The calculations predict a 1$\mu_B$ magnetic moment
for this center in agreement with earlier
LDA+$U$ studies~\cite{PhysRevB.87.144422,PhysRevB.90.125130}, but in disagreement
with the result of HSE06 calculations~\cite{ApplPhysLett.100.252904} in which
the Co$^{4+}_B$ ion had a 3$\mu_B$ magnetic moment. The Co$^{4+}_B$--O distance
in our case is 1.897~{\AA}.

For an isolated Co$^{2+}_B$--nearest~$V_{\rm O}$ complex, our calculations predict
the magnetic moment of 1$\mu_B$, in agreement with Ref.~\cite{PhysRevB.87.144422}
and in disagreement with Ref.~\cite{PhysRevB.90.125130}.

The impurity centers with trivalent cobalt at the $B$~site suggest the existence
of defects that compensate for the difference in ionic charges. Calculations for
a model, in which an additional electron was supplied to an isolated Co~ion by
a distant donor F~atom at the oxygen site (Fig.~\ref{fig3}(b), Table~\ref{table1}),
showed that the lowest-energy structure is the low-spin state ($S = 0)$, whereas
the high-spin state ($S = 2$) is characterized by a much higher energy. It is
interesting that another approach in which one extra electron was added to the
system containing an isolated Co$^{4+}_B$ ion yielded a rather close result (see
Table~\ref{table1}).

The finding that the isolated Co$^{3+}_B$ ion has no magnetic moment contradicts
the experimental data. An effective magnetic moment of Co atoms of 1.96--2.85$\mu_B$
was obtained from the temperature dependence of magnetic susceptibility in
bulk samples with Co concentration close to that in our
samples~\cite{InorgChem.43.8169,JMagnMagnMater.305.6,JMagnMagnMater.321.3526}.
Therefore, we should consider Co$^{3+}_B$ impurity complexes with oxygen vacancies.

We considered an isolated Co$^{3+}_B$--nearest $V_{\rm O}$ complex in which the
required Co oxidation state is provided by a distant acceptor impurity (Rb at
the Sr site, Fig.~\ref{fig3}(c)). It turned out that this defect is
paramagnetic and has a magnetic
moment of 2$\mu_B$ ($S = 1$). The intermediate-spin state of this complex
qualitatively agrees with the result obtained in Ref.~\cite{ApplPhysLett.100.252904}.
The effective magnetic moment for a center with $S = 1$ in the case of totally
quenched orbital momentum is $2\sqrt{S(S + 1)}\mu_B \approx 2.828\mu_B$ and is
close to the moment observed in the experiment.

\begin{table*}
\caption{\label{table2}Energies of different complexes containing two Co$_B^{3+}$
ions and one oxygen vacancy. Atomic coordinates of Co atoms and the vacancy
are given in Table~S2 of the Appendix.}
\begin{ruledtabular}
\begin{tabular}{cccc}
Configuration & Figure      & Spins of                  & Energy \\
              &             & Co(1) and Co(2) ions      & (meV/Co) \\
\hline
No nearest Co atoms, distant $V_{\rm O}$        & Fig.~3(e) & 0; 0 &  0 \\
The same (another configuration)                & Fig.~3(f) & 0; 0 & 39 \\
Nearest Co atoms, distant $V_{\rm O}$           & Fig.~3(g) & 0; 0 & 237 \\
Nearest Co atoms, nearest $V_{\rm O}$ (angular) & Fig.~3(h) & 0; 1 & 236 \\
Nearest Co atoms, nearest $V_{\rm O}$ (linear)  & Fig.~3(i) & 0; 1 & 313 \\
Nearest Co atoms, $V_{\rm O}$ at the midpoint (FM)  & Fig.~3(d) & 1; 1 & 568 \\
Nearest Co atoms, $V_{\rm O}$ at the midpoint (AFM) & Fig.~3(d) & 1; 1 & 585 \\
\end{tabular}
\end{ruledtabular}
\end{table*}

The calculations of energies of different complexes containing
two Co$^{3+}_B$ ions and one oxygen vacancy (Fig.~\ref{fig3}(d--i),
Table~\ref{table2}) brought us to the following conclusions. (1)~To have a
non-zero magnetic moment, the Co$^{3+}_B$ ion should have a nearest vacancy
(isolated Co$^{3+}_B$ ions and oxygen vacancies have zero magnetic moment);
(2)~the total energy of a system
with two nearest Co atoms is higher than that for a system with two distant Co
atoms (there is no tendency to clustering, in agreement with Ref.~\cite{PhysRevB.87.144422});
(3)~the total energy of the Co$^{3+}_B$--Co$^{3+}_B$--nearest $V_{\rm O}$ complex
depends on its configuration: the energy of the angular configuration is by
76~meV lower than the energy of the linear one. We note that the energy of the
angular configuration with the nearest oxygen vacancy is by 1~meV lower than
the energy of a pair of neighboring Co$^{3+}_B$ ions with a distant vacancy.
This result disagrees with the results reported in Ref.~\cite{PhysRevB.87.144422,
PhysRevB.90.125130} for the Co$^{2+}_B$--$V_{\rm O}$ complexes; (4)~the
ferromagnetic coupling energy of two Co$^{3+}_B$ ions with $S = 1$ in the
complex, in which the vacancy is located at the midpoint between two nearest
Co atoms, is $2J = 33$~meV.

All distances obtained in the models in which the Co$^{3+}$ ion enters the
$B$~site (Table~\ref{table1}) are in a reasonable agreement with the distance
(1.909~{\AA}) obtained in the experiment. Taking into account that the experiment
indicates the existence of magnetic moments on Co$^{3+}$ ions, we should modify
the model that was used in the EXAFS data analysis in Sec.~\ref{sec3} and consider
a model in which cobalt forms a complex with the nearest oxygen vacancy.

\section{Refinement of structural models and discussion}

Insufficiently good agreement between the experimental and calculated EXAFS
spectra stimulated us to consider more complex models. First, we considered the
model in which the Co$^{3+}_B$ ion forms a complex with the nearest oxygen
vacancy. In this model, the distances obtained for the Co3B1100 sample were
not much different from those for the model of Co$^{3+}_B$ ion with a distant
vacancy considered in Sec.~\ref{sec3}, but the $S_0^2$ factor in this model
became much closer to that in the reference compounds.

Using the results of first-principles calculations of the local environment
for the off-center
Co$^{2+}_A$ ion (Sec.~\ref{sec4}), we reanalyzed the EXAFS spectrum of the
Co2A1600 sample. It appeared that the configurations with atomic displacements
along the [100] and [111] axes give the lowest values of $R$~factor. The
calculated Fourier transform of the EXAFS spectrum for the model with [110]
displacement exhibited a strong peak resulting from the Ti atoms in the third
shell, which is absent in the experimental EXAFS spectrum. This means that this
model can be excluded from further analysis. Taking into account the significant
energy difference between the configurations with [100] and [111] displacements,
we believe that Co atoms displace along the [100] axis, and will use this model
in the future analysis.

\begin{figure*}
\includegraphics[scale=1.25]{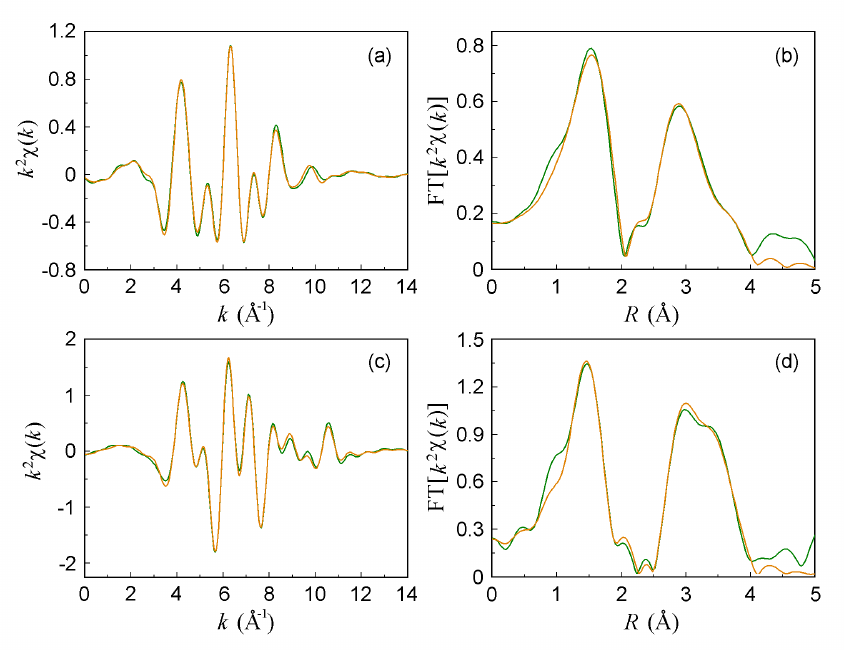}
\caption{\label{fig4}Comparison of the experimental (green) and calculated (orange)
$k^2\chi(k)$ curves in (a,c) $k$ and (b,d) $R$ spaces for (a,b) 2A1600 and (c,d)
3B1100 samples.}
\end{figure*}

The determination of the structural parameters for a two-component model in
which the EXAFS spectrum for each sample is composed of two above-considered
contributions, which act simultaneously, was performed using an iterative
technique described in detail in the Appendix. The experimental EXAFS
spectra and their best fits for the Co2A1600 and Co3B1100 samples are shown in
Fig.~\ref{fig4}. The obtained parameters for these samples are given in Table~S1
of the Appendix. For the Co$^{3+}_B$ ion, the Co--O distance in
the first shell (1.906(11)~{\AA}) and the distances to the second and third
shells are in good agreement with the results of first-principles calculations
(Table~\ref{table1}). For the Co$^{2+}_A$ ion displaced from the site in the
[100] direction, the Co--O distance in the first shell (2.040(6)~{\AA}) and
the distances to other shells are also in qualitative agreement with the
calculated data. The data analysis showed that in the Co2A1600 sample
76\% of the incorporated cobalt is at the $A$~site, whereas in the Co3B1100
sample only 18\% of impurity atoms enter the $A$~site.

The same iterative procedure was applied to the Co2A1500 sample. The
structural parameters for this sample were close to those obtained for the
Co2A1600 sample, but the fraction of Co atoms that enter the $A$~site was 75\%.

We note that the behavior of cobalt in SrTiO$_3$ is very different from that of
its neighbor in the Periodic Table~--- nickel~\cite{PhysSolidState.56.449,PhysSolidState.59.1512}.
First, only for cobalt its appreciable amount can be incorporated into the
$A$~site. Second, the magnetic properties of the impurities at the $B$~site
are exactly the opposite: for Ni, a non-zero magnetic moment appears only in
the presence of a distant oxygen vacancy, whereas for Co it appears only in
the case of the nearest vacancy.

\begin{figure*}
\includegraphics[scale=1.25]{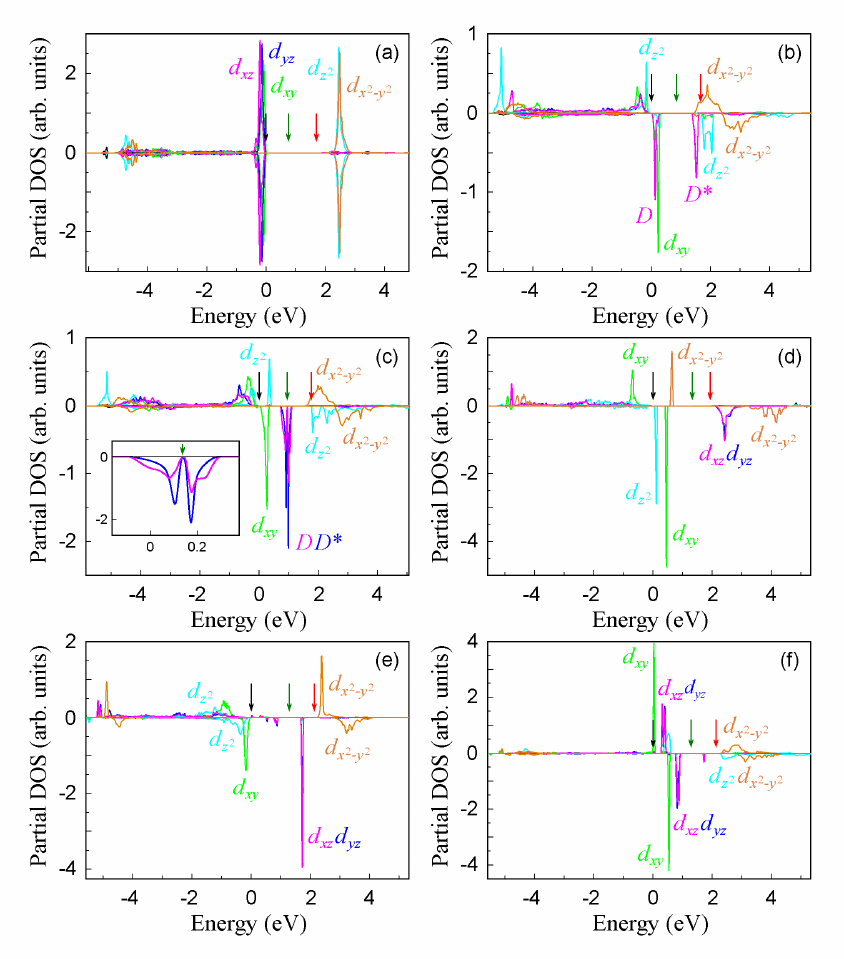}
\caption{\label{fig5}Orbitally-resolved local density-of-states of the Co atom for the
following models: (a) isolated Co$^{3+}_B$ ion, (b) Co$^{3+}_B$--nearest~$V_{\rm O}$
complex, (c) Co$^{3+}_B$--$V_{\rm O}$--Co$^{3+}_B$ complex,
(d) off-center Co$^{2+}_A$ ion, (e) off-center Co$^{3+}_A$ ion,
(f) Co$^{2+}_B$--$V_{\rm O}$ complex. The positions of the top of the valence
band, the bottom of the conduction band, and the Fermi level are shown by
black, red, and green arrows, respectively.}
\end{figure*}

The electronic structure of all considered Co impurity centers in SrTiO$_3$ is
shown in Fig.~\ref{fig5}. The isolated diamagnetic Co$^{3+}_B$ ion
(Fig.~\ref{fig3}(b)) exhibits a classical $t_{2g}$--$e_g$ splitting induced by
the octahedral crystal field (Fig.~\ref{fig5}(a)); its symmetry is slightly lowered because of the
presence of a distant F~atom. The electronic levels formed by $t_{2g}$
orbitals are merged with the valence band (this decreases the band gap by
$\sim$0.16~eV) and the $e_g$ orbitals form resonant energy levels deep in
the conduction band (the conduction band is formed from $d$~orbitals of Ti).
The Fermi level lies in the forbidden gap.

The electronic structure of the paramagnetic Co$^{3+}_B$--nearest $V_{\rm O}$
complex (Fig.~\ref{fig3}(c)) exhibits an interesting feature. In addition
to the energy level in the forbidden gap, which lies close to the valence band
and is formed from the $d_{xy}$ orbital of Co, there appears a pair of energy
levels $D$ and $D^*$ formed from $d_{xz}$ and $d_{yz}$ orbitals (Fig.~\ref{fig5}(b)).
One of these levels is occupied and the other one is unoccupied by electron. Both
levels are located in the forbidden gap, and the Fermi level is located between
them. Our explanation is that the neighboring vacancy distorts the crystal field
so strong that these orbitals transform to two very distinct wave functions, which
have a tetragonal symmetry and so, when calculating the projections onto the
$Y_{2,-1}$ and $Y_{2,1}$ spherical harmonics, give the same coefficients (magenta
and blue lines on the figure coincide). There is one more broad impurity band formed
from $d_{x^2-y^2}$ orbital, which is partially superimposed on the conduction
band and forms a tail in the forbidden gap. The spin-minority $d_{z^2}$ band is
superimposed on the conduction band.

The electronic structure of the paramagnetic Co$^{3+}_B$--$V_{\rm O}$--Co$^{3+}_B$
complex (Fig.~\ref{fig3}(d)) reminds that of the above
Co$^{3+}_B$--$V_{\rm O}$ complex. In the forbidden gap it also has a pair of
energy levels formed from $d_{xz}$ and $d_{yz}$ orbitals (Fig.~\ref{fig5}(c)),
one of which ($D$) is
occupied and the other one ($D^*$) is unoccupied by electron, but in this case
the energy gap between them is small ($\sim$60~meV, see insert in Fig.~\ref{fig5}(c)).

The paramagnetic Co$^{2+}_A$ ion (Fig.~\ref{fig3}(a)) creates three energy
levels in the lower half of the forbidden band of SrTiO$_3$ (Fig.~\ref{fig5}(d)).
The lowest of them, formed from the spin-down $d_{z^2}$ orbital, merges with
the valence band, whereas two other levels, formed from spin-down $d_{xy}$
and spin-up $d_{x^2-y^2}$
orbitals, are filled with electrons. Three resonant levels are located
deep in the conduction band. The Fermi level lies in the forbidden gap.

An interesting effect is observed when studying the configuration containing
the Co$^{2+}_A$ ion and the Co$^{3+}_B$--$V_{\rm O}$ complex simultaneously.
It turned out that the equilibrium charge states of these centers are changed
because of the transfer of one electron from the Co$^{2+}_A$ ion to the
Co$^{3+}_B$--$V_{\rm O}$ complex. This transfer is accompanied by a large
reconstruction of the electronic structure of both centers. After the
electron from the spin-up $d_{x^2-y^2}$ orbital of the Co$_A^{2+}$ ion is
removed, two occupied $d$ orbitals shift downward (compare
Figs.~\ref{fig5}(d) and \ref{fig5}(e)) thus decreasing the band-structure
energy by $\sim$1~eV. On the contrary, the Co$^{3+}_B$--$V_{\rm O}$ complex,
when accepting one electron, captures this electron on the unoccupied $D^*$
orbital. This triggers a large reconstruction of orbitals including the
removing of the $D$--$D^*$ splitting, downward shift of this pair of orbitals,
and upward shift of all other occupied orbitals (compare Figs.~\ref{fig5}(b)
and \ref{fig5}(f)). The magnetic moments of the Co$^{3+}_A$ ion and the
Co$^{2+}_B$--$V_{\rm O}$ complex become $2\mu_B$ and $1\mu_B$, respectively.

One can expect that under the interband optical excitation, the Co$^{3+}_A$ ion
and the Co$^{2+}_B$--$V_{\rm O}$ complex can be transformed back to the Co$^{2+}_A$
and Co$^{3+}_B$--$V_{\rm O}$ centers as a result of capturing of the light-excited
electrons and holes on the oppositely charged centers. As the magnetic moments
of these centers increase to $3\mu_B$ and $2\mu_B$, we can anticipate an
interesting effect: optically-induced changes of magnetic properties of Co-doped
SrTiO$_3$. As far as we know, this effect was not studied so far. The
possibility of this effect is confirmed by photochromism (optically-induced
reversible changes in optical absorption spectra) observed earlier in Co-doped
SrTiO$_3$~\cite{PhysRevB.4.3623}.

Based on our previous studies of Ni-doped SrTiO$_3$ and
BaTiO$_3$~\cite{PhysSolidState.56.449,PhysSolidState.59.1512}, we deliberately
used low concentration of Co in our calculations to avoid the appearance of
broad impurity bands. First-principles calculations of
Ref.~\cite{CommunPhys.23.263} showed that Co forms an impurity band in the
lower half of the forbidden band, and at $x = 0.25$ its width becomes so large
that the band structure characteristic of metals is formed. Experimental data
confirm this prediction: SrTiO$_3$(Co) samples with $x > 0.2$ completely absorb
light in the visible and infrared regions~\cite{CommunPhys.23.263}.

In Ref.~\cite{MaterResBull.37.1215}, the acceptor character of the cobalt
impurity with an activation energy of $E_a = 1.3$~eV was established from the
studies of conductivity of Y-codoped SrTiO$_3$(Co) samples under variable partial
oxygen pressure. According to our calculations, the Fermi level in all studied
Co impurity centers lies close to the center of the forbidden band gap
(Fig.~\ref{fig5}). The above calculations for the model where two centers are
simultaneously present in the sample showed that the Co$^{2+}_A$ can donate one
electron, whereas the Co$^{3+}_B$--$V_{\rm O}$ complex can accept it. Therefore,
the properties of the Co$^{3+}_B$--$V_{\rm O}$ complex which is present in
all samples according to our EXAFS data are consistent with the observed
electrical properties.

In considering optical transitions involving local energy levels in the
forbidden gap, we should concentrate on the $p$--$d$ transitions between the
valence band (which is formed primarily from the O $2p$ states) and unoccupied
$d$ states. The intracenter $d$--$d$ transitions and transitions between
occupied $d$ levels in the forbidden gap and the conduction band (which is
formed primarily from the Ti $3d$ states) are always weaker because the initial
and final states have the same parity. Our electronic structure calculations
(Fig.~\ref{fig5}) show that for the Co$^{3+}_B$--$V_{\rm O}$ complexes the
$p$--$d$ transitions to the $D^*$ states can result in strong absorption,
whereas for the Co$^{2+}_A$ and isolated Co$^{3+}_B$ ions such transitions are
not possible. On the contrary, both the Co$^{3+}_A$ ion and the
Co$^{2+}_B$--$V_{\rm O}$ complex exhibit unoccupied $d_{xz}$ and $d_{yz}$ levels,
which can produce strong optical absorption. The appearance of these levels
in the forbidden gap can explain the dark brown color of Co-doped SrTiO$_3$
samples and suggest a possible application of this material in solar-light-driven
photocatalytic converters. The SrTiO$_3$(Co) samples has already demonstrated
high catalytic activity in the oxidation of propane, methane, and carbon
monoxide~\cite{MaterResBull.37.1215}.

In summary, the Co$^{3+}_B$--$V_{\rm O}$ and Co$^{2+}_B$--$V_{\rm O}$ complexes
are probably the main defects whose properties are consistent with magnetic,
optical, electrical, and structural data obtained from the experiment. Other
defects considered in this work can also affect these properties.

We would like to draw attention to the fact that the preparation conditions of
the SrTiO$_3$(Co) samples in which the room-temperature ferromagnetism was
observed are far from equilibrium. We can suppose that a significant amount of
the off-center Co$^{2+}_A$ ions which have a large magnetic moment (3$\mu_B$)
can appear in these samples in non-equilibrium conditions. An increase in
the lattice parameter experimentally observed in these samples~\cite{NewJPhys.12.043044}
support this idea: the Co$^{2+}_A$ center is the only structural defect
we have modeled which increased the unit cell volume upon doping. We think
the question about the local structure and oxidation state of structural defects
in these samples needs further experimental investigation.

\section{Conclusions}

XAFS technique combined with first-principles calculations has been demonstrated
to be a very fruitful approach for studying the structure of impurity-induced
defects in crystals. In this work, the local structure and the oxidation state
of the Co impurity in ceramic SrTiO$_3$ have been studied by XAFS spectroscopy.
The change in the synthesis
conditions was shown to significantly influence the cobalt concentrations at the
$A$ and $B$~sites of the perovskite structure. At an annealing temperature of
1600$^\circ$C, up to 76\% of Co atoms enter the $A$~site. The Co ions at the
$A$~site are divalent, whereas the ions entering the $B$~site are trivalent.
It was revealed that the Co$^{2+}_A$ ions are off-center and are displaced from
the $A$~site by 1.0~{\AA}.

First-principles calculations of the properties of different structural defects
in Co-doped SrTiO$_3$ were used to determine the defects whose oxidation state,
local environment, magnetic, electrical, and optical properties agree with
experiment. In particular, it was shown that the off-center Co$^{2+}_A$ ions
have a magnetic moment of $3\mu_B$. As concerns to Co at the $B$~site, only
the impurity--nearest vacancy complexes have a non-zero magnetic moment.
Electronic structure calculations of different Co-containing defects show
that the Co$^{3+}_B$--$V_{\rm O}$ complexes as well as Co$^{3+}_A$ ion and the
Co$^{2+}_B$--$V_{\rm O}$ complex can produce strong optical absorption which can
explain the brown color of Co-doped samples.

A new effect, the optically-induced change of magnetic properties, was predicted
for samples containing Co$^{2+}_A$ and Co$^{3+}_B$--$V_{\rm O}$ defects
simultaneously.

And the last, the reorientation of dipole moments associated with off-center
Co$^{2+}_A$ ions by an external electric field can strongly affect the
interaction of their magnetic moments with other magnetic Co$^{3+}_B$ complexes,
thus inducing magnetoelectric effect in the samples.

\begin{acknowledgments}
This work was supported by the Russian Foundation for Basic Research (Grant
No.~17-02-01068). The authors are grateful to the BESSY staff for hospitality
and financial support during their stay at the laboratory.
\end{acknowledgments}

\appendix*

\section{}

\addtocounter{table}{-2}
\renewcommand{\thetable}{S\arabic{table}}

The determination of the structural parameters for a two-component model in which
the EXAFS spectrum for each sample is composed of two independent contributions,
which will be called the $A$ and $B$~states according to the lattice sites of
strontium titanate into
which the cobalt atoms enter, was performed by an iterative technique. As
the starting point, it was assumed that the spectrum for the $B$~state is the
spectrum of the Co3B1100 sample. For this spectrum, a set of structural parameters
(distances and Debye-Waller factors for three shells) was obtained by fitting.
After that, the spectrum of the Co2A1600 sample (and then of the Co2A1500 one) was
represented as a sum of the $A$ and $B$~states with unknown proportion. Having
fixed the parameters for the $B$~state determined on the previous step and taking
the interatomic distances calculated for the off-center Co$^{2+}_A$ ion in Sec.~IV
of the main paper as the second starting point, the structural parameters for the
$A$~state and the relative contributions of $A$ and $B$~states to the spectrum
were calculated. As the next step, the spectrum for the Co3B1100 sample
was re-processed, assuming that it may contain an unknown fraction of the $A$~state.
This fitting was performed with fixed parameters determined on the previous step
for the $A$~state. This gave us the refined parameters for
the $B$~state and the relative amounts of the $A$ and $B$~states in the spectrum.
The multiple repetition of the described procedure made it possible to obtain
the refined structural parameters for the $A$ and $B$~states and their relative
contributions to the analyzed spectra.

The set of parameters obtained for Co2A1600 and Co3B1100 samples are given in
Table~\ref{tableS1}.

Table~\ref{tableS2} gives the positions of two Co atoms and the oxygen vacancy
used in modeling of the electronic structure and the corresponding figures in the
main paper.

\onecolumngrid

\begin{table*}
\caption{\label{tableS1}Structural parameters obtained from the EXAFS data
analysis of two studied samples.}
\begin{ruledtabular}
\begin{tabular}{ccccccc}
Sample  & Site  & $R$~factor & $S_0^2$ & Shell & $R_i$~({\AA}) & $\sigma_i^2$ ({\AA}$^2$) \\
\hline
Sr$_{0.98}$Co$_{0.02}$O$_3$ & $A$ & 0.00297 & 0.705 & Co--O  & 2.040(6)  & 0.0049(18) \\
annealed at 1600$^\circ$C   &     &         &       & Co--Ti & 3.094(15) & 0.0146(27) \\
                            &     &         &       & Co--Sr & 3.816(20) & 0.0114(40) \\
                            & $B$ &         & 0.282 & Co--O  & 1.906(11) & 0.0037(18) \\
                            &     &         &       & Co--Sr & 3.354(19) & 0.0087(12) \\
                            &     &         &       & Co--Ti & 3.907(17) & 0.0064(13) \\
\hline
SrTi$_{0.97}$Co$_{0.03}$O$_3$ & $A$ & 0.00255 & 0.153 & Co--O  & 2.040(6)  & 0.0049(18) \\
annealed at 1100$^\circ$C     &     &         &       & Co--Ti & 3.094(15) & 0.0146(27) \\
                              &     &         &       & Co--Sr & 3.816(20) & 0.0114(40) \\
                              & $B$ &         & 0.731 & Co--O  & 1.906(11) & 0.0037(18) \\
                              &     &         &       & Co--Sr & 3.354(19) & 0.0087(12) \\
                              &     &         &       & Co--Ti & 3.907(17) & 0.0064(13) \\
\end{tabular}
\end{ruledtabular}
\end{table*}

\begin{table*}
\caption{\label{tableS2}Configurations of different complexes containing two Co$_B^{3+}$
ions and one oxygen vacancy.}
\begin{ruledtabular}
\begin{tabular}{ccccc}
Configuration & \multicolumn{3}{c}{Positions of two Co atoms and vacancy} & Figure in the \\
\cline{2-4}
              & Co(1) & Co(2) & $V_{\rm O}$                               & main paper \\
\hline
No nearest Co atoms, distant $V_{\rm O}$        & (0,0,0) & (0.5,0.5,0) & (0.5,0.5,0.5) & Fig.~3(e) \\
The same (another configuration)                & (0,0,0) & (0.5,0.5,0.333) & (0,0,0.5) & Fig.~3(f) \\
Nearest Co atoms, distant $V_{\rm O}$           & (0,0,0) & (0,0,0.333) & (0.5,0.5,0.5) & Fig.~3(g) \\
Nearest Co atoms, nearest $V_{\rm O}$ (angular) & (0,0,0) & (0,0,0.333) & (0, 0.25,0)   & Fig.~3(h) \\
Nearest Co atoms, nearest $V_{\rm O}$ (linear)  & (0,0,0) & (0,0,0.333) & (0,0,0.833)   & Fig.~3(i) \\
Nearest Co atoms, $V_{\rm O}$ at the midpoint (FM)  & (0,0,0) & (0,0,0.333) & (0,0,0.167) & Fig.~3(d) \\
Nearest Co atoms, $V_{\rm O}$ at the midpoint (AFM) & (0,0,0) & (0,0,0.333) & (0,0,0.167) & Fig.~3(d) \\
\end{tabular}
\end{ruledtabular}
\end{table*}

\twocolumngrid

\providecommand{\BIBYu}{Yu}

\end{document}